\documentclass[fleqn,usenatbib]{mnras}
\usepackage{newtxtext,newtxmath}

\usepackage[T1]{fontenc}

\DeclareRobustCommand{\VAN}[3]{#2}
\let\VANthebibliography\thebibliography
\def\thebibliography{\DeclareRobustCommand{\VAN}[3]{##3}\VANthebibliography}

\usepackage{graphicx}	
\usepackage{amsmath}	
\usepackage{natbib}
\usepackage{ulem}
\usepackage{CJKutf8}    
\usepackage{alphabeta}

\title[H{\tt I} absorption of PSR B0458+46]{Distance of PSR B0458+46 indicated by {\it FAST} H{\tt I} absorption observations}

\author[Jing et al.]{
W. C. Jing,$^{1,2}$
J. L. Han,$^{1,2}$\thanks{E-mail: hjl@nao.cas.cn (JLH)}
Tao Hong,$^{1}$ 
Chen Wang,$^{1,2}$ 
X. Y. Gao,$^{1,2}$ 
L. G. Hou,$^{1,2}$
D. J. Zhou,$^{1,2}$
\newauthor
J. Xu,$^{1}$
Z. L. Yang$^{1,2}$ 
\\
\\
$^{1}$National Astronomical Observatories, Chinese Academy of Sciences, 20A Datun Road, Chaoyang District, Beijing 100101, China\\
$^{2}$School of Astronomy, University of Chinese Academy of Sciences, Beijing 100049, China\\
}

\date{Accepted XXX. Received YYY; in original form ZZZ}

\pubyear{2022}

\begin{document}
\label{firstpage}
\pagerange{\pageref{firstpage}--\pageref{lastpage}}
\maketitle

\begin{abstract}
The pulsar B0458+46 was previously believed to have a distance of about 1.3~kpc and to be associated with a nearby supernova remnant, SNR HB9 (G160.9+2.6). We observe the neutral hydrogen (H{\tt I}) absorption spectrum of PSR B0458+46  by using the \textit{Five-hundred-meter Aperture Spherical radio Telescope (FAST)}, and detect two absorption lines at radial velocities of $V_{\rm LSR} = {-7.7}~{\rm km~s}^{-1}$ and {$-28.1$}~km~s$^{-1}$. Based on the Galactic rotation curve with a modification factor correcting for the systematic stream in the Galactic anti-center region, we derive the kinematic distance of the farther absorption cloud, which is found to be located $2.7^{+0.9}_{-0.8}$ kpc away, just beyond the Perseus Arm. We also obtain a direct distance estimation of the farther absorption cloud as being $2.3_{-0.7}^{+1.1}$ kpc, based on a comparison of the velocity with the H{\tt I} emission in the Perseus and Outer Arms that was well-defined by recently measured parallax tracers. As a result, we conclude that PSR B0458+46 should be located beyond the Perseus Arm, with a lower limit for the distance at 2.7 kpc, and therefore not associated with SNR HB9. The doubled distance indicates a deficiency of thermal electrons in the immediate outer Galaxy, with a much less density than current models predict. Additionally, we detect a new high-velocity H{\tt I} cloud in the direction of this pulsar.
\end{abstract}



\section{Introduction}

Pulsar distance is a fundamental parameter for several related studies such as population syntheses \citep[e.g. ][]{lfl06} and pulsar luminosity function \citep[e.g. ][]{wzz20}. The distance to a radio pulsar can be inferred from the dispersion measure (DM) by ${DM} = \int_0^{\rm distance}{n_{\rm e}~{{\rm d}s}}$ based on a distribution model for electron density $n_{\rm e}$, with ${\rm d}s$ being the unit segment along the line of sight. 
However, independent measurements of pulsar distances are necessary to construct a reliable model of the Galactic free electron density distribution \citep[e.g.][]{tc93,cl02,ymw17}. 
There are three methods for obtaining an independent pulsar distance \citep{fw90,vwc+12}: (1) direct parallax measurements of pulsars \citep[e.g.][]{dgb19}, (2) association with an object of known distance \citep[e.g.][]{csl02,lsg09}, and (3) kinematic distance of low-latitude pulsars derived from foreground neutral hydrogen (H{\tt I}) clouds \citep[e.g.][]{wbr79,lks15}.

Kinematic distance constraints can be obtained from the absorption of pulsar emission by foreground H{\tt I} clouds \citep{fw90}. The distance of an H{\tt I} cloud located in the Galactic plane can be estimated through a Galactic rotation curve model that converts the measured radial velocity of the cloud into distance \citep[e.g.][]{bs65,fbs89,bb93,rmb19}. If foreground H{\tt I} clouds obscure a pulsar, the H{\tt I} absorption lines should appear against the emission of that pulsar \citep[e.g.][]{wbr79,wrb80,wsx+08,fw90}, which sets a lower limit on the pulsar distance. Conversely, if a cloud does not produce any absorption, the measurements of the cloud could provide an upper limit for the pulsar distance. The kinematic distance constraints of 62 pulsars have been utilized to model the distribution of free electrons in the Galaxy \citep[e.g.][]{ymw17}.

To successfully observe and resolve the H{\tt I} absorption line in pulsar emission, either the pulsar must be strong or the radio telescope must be sensitive enough to achieve a reasonable signal-to-noise (S/N) ratio within an affordable observation time \citep[see, e.g.][]{fw90}. The {\it Five-hundred-meter Aperture Spherical radio Telescope} \citep[{\it FAST},][]{Nan2006,Nan2011} is a highly sensitive radio telescope capable of efficiently observing H{\tt I} absorption spectra better than other telescopes \citep[e.g.][]{yff22}.

PSR B0458+46 (J0502+4654) is a pulsar that was first discovered by \citet{dth78} during the early scanning survey of the old 300-feet (91~m) transit radio telescope at Green Bank. It is located in the Galactic disk at the Galactic coordinates of $(l,b) = (160.3628^{\circ},+3.0766^{\circ})$. PSR B0458+46 has a spin period of 0.6386~s and a $\rm DM$ of $41.834$~pc~cm$^{-3}$ \citep[see the updated webpage\footnote{\url{https://www.atnf.csiro.au/people/pulsar/psrcat/}} version of][]{mht05}. Its distance is estimated to be about 1.3 kpc according to its DM value using the Galactic electron density model \citep[e.g.][]{cl02,ymw17}. PSR B0458+46 is located inside the supernova remnant (SNR) HB9 (G160.9+2.6) in the sky, raising suspicions that the two objects might be physically associated \citep[e.g.][]{dth78,mwk03,lt07}. This SNR is probably very nearby, located within 1~kpc \citep{lt07,zjl20}. However, there have been no independent distance measurements for this pulsar.

We use the {\it FAST} to observe the H{\tt I} absorption line of PSR B0458+46. The observations and data processing methods for obtaining the pulsar H{\tt I} absorption spectrum are described in Section 2. In Section 3, we present the results of our {\it FAST} observations. In Section 4, we discuss relevant issues, including the potential association with the SNR HB9 and the electron density distribution in the outer Galactic disk. We summarize our findings in Section 5. As a by-product, we detect a new high-velocity H{\tt I} cloud in the line of sight of this pulsar, which is presented in the Appendix \ref{app:hi}.

\begin{table}
    \caption{FAST observation parameters for PSR B0458+46.}
    \label{tab:obs_para}
    \begin{tabular}{ll}
      \hline\hline
      Observational length (s) @ Date & 1260 @ 2022.08.18  \\
      & 1260 @ 2022.08.22 \\
      & 3600 @ 2022.09.17 \\[1mm] 
      {\it {\it FAST} spectral backend:} &\\
      Spectral sampling time (s) & 0.1006633 \\
      Spectral central frequency (MHz) & 1420 \\
      Spectral bandwidth (MHz) & 31.25 \\
      Spectral frequency channel number  & 65536 \\
      Spectral frequency resolution (kHz)  & 0.4768 \\ 
      Spectral velocity resolution ($\rm km~s^{-1}$)  & 0.1007  \\
      Spectral polarization products  & XX, YY, X$^*$Y and XY$^*$ \\ [5mm]
      {\it {\it FAST} pulsar backend:} &\\
      PSR sampling time (\mu\!s)  & 49.152  \\
      PSR frequency range (MHz) & 1000 -- 1500 \\
      PSR frequency channel number & 2048 \\
      PSR polarization products  & XX, YY, X$^*$Y and XY$^*$ \\
      \hline
    \end{tabular}
\end{table} 

\section{Observations and Data Reduction}

\subsection{Observations}

We conducted {\it FAST} observations of PSR B0458+46 using the central beam of the L-band 19-beam receiver \citep{jth20} on August 18 and 22 as the main part of project PT2022\_0197, and on September 17 in the context of project ZD2022\_2. The observational parameters are summarized in Table~\ref{tab:obs_para}. Both the spectral backend and pulsar backend were used to simultaneously record data. The receiver covers a frequency range of 1000 -- 1500~MHz. At the beginning of every observation session, the calibration signals with an amplitude of $T_{\rm noise} = 1.1$~K and an on-off period of 2.01326~s are injected to the receiver feed for 2~min, and the data with these calibration signals are recorded for calibrations.

To observe the H{\tt I} line, we used the N mode of the {\it FAST} spectral backend, which splits signals from a bandwidth of 31.25~MHz around 1420~MHz into 65536 frequency channels. This provides a frequency resolution of 0.4768~kHz, corresponding to a velocity resolution of 0.1~km$\rm ~s^{-1}$ for the H{\tt I} line. The spectral backend accumulates and saves data every 0.1~s during the observations. 

During all observation sessions, we use the {\it FAST} pulsar backend to simultaneously record searching-mode data with a time resolution of 49.152 \mu\!s for 2048 frequency channels that cover the frequency range of 1000 -- 1500~MHz, each with the 4 polarization products (XX, YY, X$^*$Y and XY$^*$). 

\begin{figure}
  \centering
  \includegraphics[width=0.95\columnwidth]{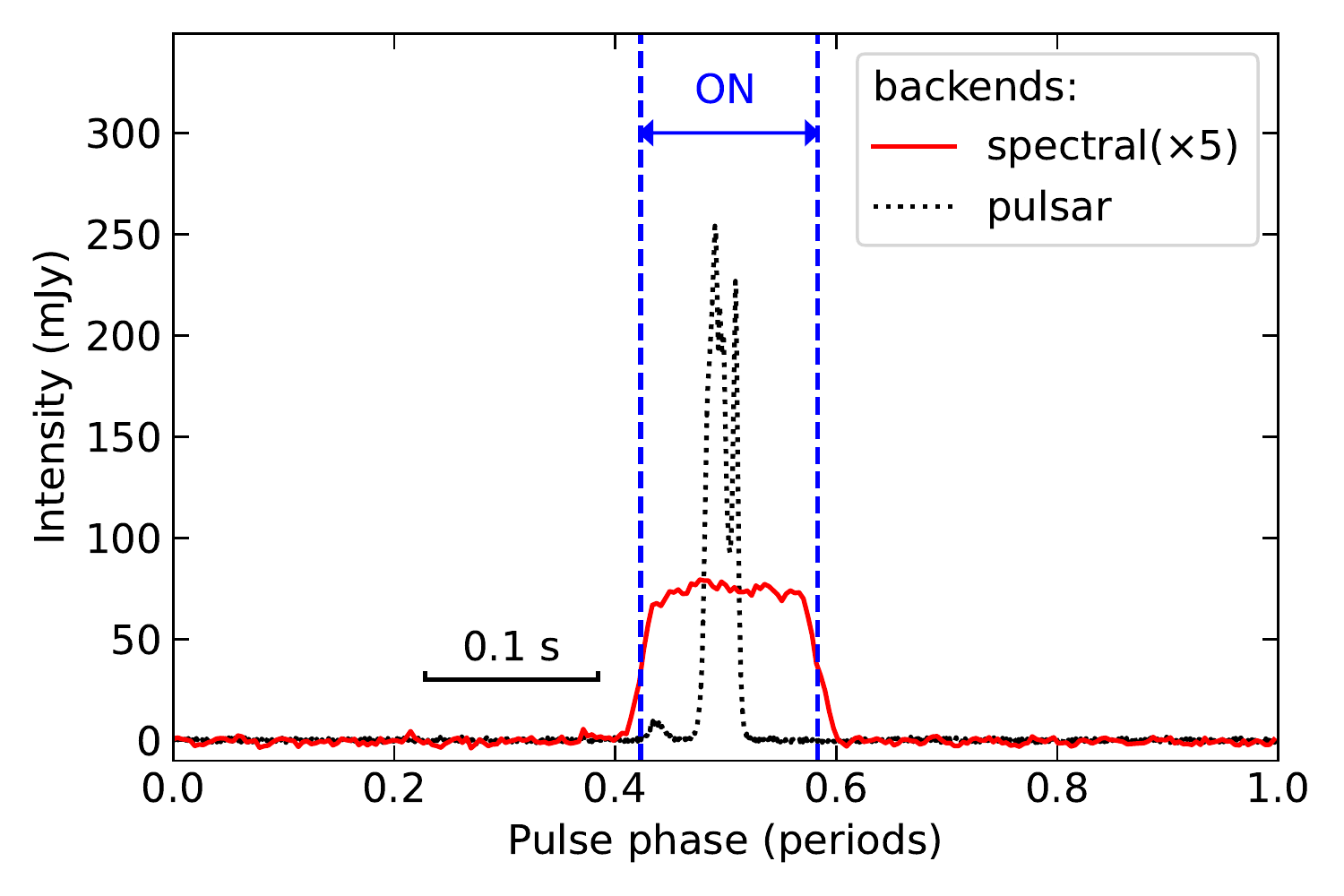}
  \caption{The integrated pulse profiles of PSR B0458+46 obtained from data recorded by the {\it FAST} spectral backend (solid line, 256 phase bins per period) and pulsar backend (dot line, 512 phase bins per period). Note that the spectral backend saves data every 0.1~s, equivalently a sample for about 40 phase bins for the solid line. The pulse-on phase range of the folded spectral data is marked by two vertical dashed lines.}
  \label{fig:prof}
\end{figure}

\subsection{Data Processing} \label{subsec:dataprocess}

We conducted careful processing of the narrow-band spectral data. First, we manually checked the radio frequency interference (RFI) level for each observation session and found that the data around 1420 MHz were very clean, requiring no RFI mitigation for spectral data. Next, we obtained the intensity calibration factors from the 2-min sub-session with on-off calibration signals. The data were folded with a period of 2.01326~s, and the calibration factors were derived from the machine number differences between the noise-on and noise-off spectra. 
With such a scale factor, the machine number in real observation data for the total power can be converted to antenna temperature $T_{\rm A}$. A pulsar can contribute $\delta T_{\rm A}$ to the system temperature. Taking the {\it FAST} gain of $G$ = 16 K~Jy$^{-1}$ \citep[see Table 5 in][]{jth20}, one can convert the $\delta T_{\rm A}$ of pulsar to the mean flux density $S_{\nu}$ with a formula of $S_{\nu} \times P / W = \delta T_{\rm A} / G$, where $P$ is the spin period and $W$ is the pulse width.

For velocity calibration, we used the Python package \texttt{astropy} \citep{astropy13, astropy18} to correct for the Doppler effect induced by the spin and orbit of the Earth, which allowed us to measure radial velocities relative to the barycenter of the solar system. Finally, we modified the barycentric velocity of H{\tt I} clouds to the velocity $V_{\rm LSR}$ towards the local standard of rest (LSR), by adopting the standard solar motion with 20~$\rm km~s^{-1}$ towards RA (1900) of 18$^{\rm h}$ and Dec (1900) $+30^\circ$ \citep[see][]{kl86}. The values in the Galactic Cartesian frame are $(U_{\odot}, V_{\odot}, W_{\odot}) = (10.0, 15.4, 7.8)~{\rm km~s^{-1}}$. Here, $U_{\odot}$ is the velocity towards the Galactic center; $V_{\odot}$ is parallel with the circular motion of the Galaxy; and $W_{\odot}$ is towards the northern pole of the Milky Way. 

The barycentric period of PSR B0458+46 is calculated from the ephemeris of the pulsar catalog V1.67 \citep{mht05}, and produces the topocentric period for folding pulses. For each frequency channel, we fold data into 256 phase bins per pulsar period, though one sample of 0.1~s can feed into 40 phase bins. We obtain a waterfall plot in the frequency and pulse phase dimensions. 
The frequency-integrated pulse profile observed by the spectral backend is shown in Fig. \ref{fig:prof}, compared with a high time resolution pulse profile obtained by the pulsar backend folded by \texttt{DSPSR}\footnote{\url{http://dspsr.sourceforge.net/}} \citep{2011PASA...28....1V}.
The broadening of the pulse is caused by the smearing effect of the coarse sampling time of 0.1~s in the {\it FAST} spectral backend. 
According to the integrated profile in Fig. \ref{fig:prof}, the pulse-on phase is taken as the bins with an intensity greater than the half maximum, and the pulse-off range is taken as the phase range outside 0.4 -- 0.6.

\begin{figure}
  \centering
  \includegraphics[width=0.98\columnwidth]{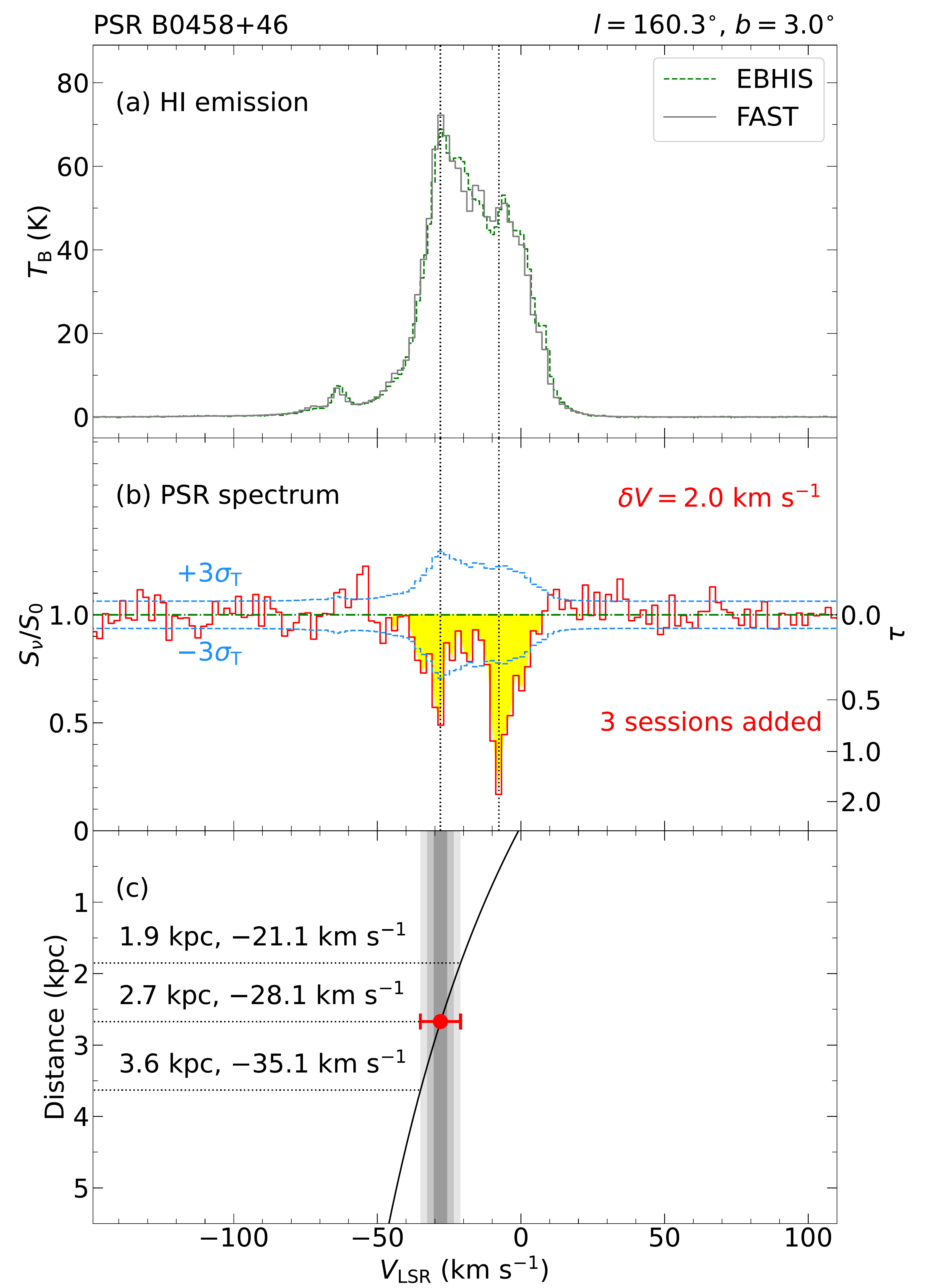}
  \caption{
  {\it FAST} observations of H{\tt I} lines for determining the lower limit for the distance of PSR B0458+46. 
  {\it (a) Top subpanel:} The observed H{\tt I} emission line integrated from all three sessions, compared with the result from the EBHIS \citep{wkf16}.
  {\it (b) Middle subpanel:} Final H{\tt I} absorption spectrum of the pulsar, obtained as the weighted sum of three sessions, plotted together with $\pm3\sigma_{\rm T}$ lines and the marked absorption line area. 
  The spectrum is scaled by its median unabsorbed value, which is scaled by {\it FAST} gain and get $S_0=2.4$~mJy. The $\sigma_{\rm T}$ lines are obtained by considering the contributions from H{\tt I} emission [see equation (\ref{equ:temp})].
  The opacity $\tau$ on the right is derived from $e^{-\tau} = S_{\nu}/S_0$.
  Two vertical dotted lines correspond to the two peaks of absorption lines.
  {\it (c) Bottom subpanel:} Velocity-distance conversion for H{\tt I} clouds on the line of sight of PSR B0458+46, that is derived from the Galactic rotation curve but modified by a factor of 1.6 (see text) caused by the systematic streaming motion in the Galactic anti-center.
  A gray shadow indicates the velocity probability range caused by the random peculiar motion of H{\tt I} clouds.}
\label{fig:all}
\end{figure}

In principle, averaging the pulse-on data and pulse-off data from the {\it FAST} spectral backend across all spectral channels can produce the pulse-on spectrum and pulse-off spectrum, respectively. However, high resolution spectral measurements can always be affected by standing waves that significantly impact spectral analyses. To obtain the pulsar absorption spectrum, we remove the standing waves naturally by subtracting the averaged pulse-off spectrum from the averaged pulse-on spectrum, which are both affected by the same standing waves. Because the spectral data are well-calibrated, we obtain an absorption spectrum with a satisfactory baseline [see Fig.~\ref{fig:all}(b) for the sum of three sessions, and Fig. \ref{fig:ghost} for 3 individual sessions]. Nevertheless, to observe the H{\tt I} emission line spectrum  [see Figure~\ref{fig:all}(a)], corrections for the standing waves and baseline in the pulse-off spectrum is necessary. We fit the data with a wave model, and obtain good results for each session (see Appendix \ref{app:hi} for details, demonstrated in Fig. \ref{fig:hi_tiny}), and finally get the H{\tt I} emission spectrum in Figures~\ref{fig:all}a.

\section{Results and discussion}

With the measurements for pulse-on and pulse-off temperature and the {\it FAST} gain of 16~K~Jy$^{-1}$, we get the flux density of PSR B0458+46 obtained from the well-calibrated different antenna temperatures between pulse-on and pulse-off spectra. PSR B0458+46 was measured by {\it FAST} to have mean flux densities of $2.3 \pm 0.3$~mJy, $2.4 \pm 0.4$~mJy and $2.5 \pm 0.2$~mJy in three observation sessions. The uncertainty of the flux densities is the root mean square (RMS) of the pulsar spectra around 1420 MHz. The averaged mean flux density at 1.42 GHz measured in the three sessions is $2.4 \pm 0.2$~mJy, which agrees with $2.5 \pm 0.1$~mJy at 1.4 GHz given by \citet{lyl95}.

The absorption spectrum of PSR B0458+46 is obtained as a weighted-average of the three measurement sessions and is shown in Fig.~\ref{fig:all}(b). Two absorption lines are detected at radial velocities of $V_{\rm LSR} = {-7.7}~{\rm km~s}^{-1}$ and {$-28.1$}~km~s$^{-1}$. Based on the radial velocity at the absorption center and the modified velocity-distance conversion curve [see Fig.~\ref{fig:all}(c) and details below], a lower limit for the pulsar distance is estimated to be 2.7 kpc. 

In the following, we discuss the results in detail.

\subsection{HI emission lines} \label{sec:hi_emission}

The H{\tt I} emission lines in the direction of PSR B0458+46 were previously obtained from available surveys, such as the Effelsberg-Bonn H{\tt I} survey \citep[i.e., EBHIS, see][]{wkf16}. Our new {\it FAST} observations have a smaller beam and a better sensitivity, but get similar H{\tt I} emission lines to the previous data. We normalize the {\it FAST} measurements to conform to the precise brightness temperature of $T_{\rm B}$ given by \citet{wkf16}. The great similarities of the line details are shown in Fig.~\ref{fig:all}(a) between our new {\it FAST} measurement and previous results.

\subsection{Pulsar H{\tt I} absorption spectrum} \label{sec:hi_absorption}

The pulsar spectrum, showing the H{\tt I} absorption line of pulsar emission as in Fig.~\ref{fig:all}(b), is the difference between the
spectra gathered in the pulse-on and pulse-off phase ranges. The pulsar spectra obtained during all three observation sessions reveal the presence of two distinct H{\tt I} absorption lines, as shown in Fig. \ref{fig:ghost} in Appendix \ref{app:psr}, probably because the H{\tt I} clouds in front of the pulsar do not exhibit significant variations \citep[e.g.][]{swh03,wsx+08,lkl21} in the time scale of only one month (rather than years). However, the low S/N ratio of each session prevents us from confidently claiming any significant changes in the observed H{\tt I} absorption lines. To improve the S/N ratio of the H{\tt I} absorption lines, especially the one with a more negative velocity which comes from distant clouds, we add these results from all three observations (Fig. \ref{fig:ghost}) and obtain the finally weighted average for the pulsar H{\tt I} absorption spectrum as shown in Fig.~\ref{fig:all}(b). No doubt that two absorption lines are detected at the velocities of $V_{\rm LSR}=-7.7$~km~s$^{-1}$ and $-28.1$~km$\rm ~s^{-1}$, with a S/N ratio of more than $4\sigma_{\rm T}$ for both. Here $\sigma_{\rm T}$ 
results from the system temperature, plus the extra noise temperature from the strong H{\tt I} emission in these frequency channels \citep{wsx+08}, see Appendix \ref{app:psr} for details.

Notice that any improper calibration of the pulse-on or pulse-off spectra may leak the strong H{\tt I} emission line into the final absorption spectrum, which then forms the  ``ghosts of H{\tt I} emission'' \citep{wrb80}. There is no evidence of this phenomenon in our {\it FAST} observation result in Fig.~\ref{fig:all}(b), probably because (1) the spectral sampling with 0.1~s enables us to resolve the \textbf{strongly fluctuating} individual pulses of PSR B0458+46, with 6 samples per period, which enable us to properly separate the phase ranges for pulse-on and pulse-off in every period; (2) both pulse-on and pulse-off spectra are well-scaled; (3) the {\it FAST} spectral data have a large dynamic range as they are expressed by 32-bit float; and (4) the baseline and standing waves are the same for both the pulse-on or pulse-off spectra and diminished properly from the spectrum subtraction.

\subsection{New distance limit of PSR B0458+46} \label{sect:dist}

\begin{figure}
  \centering
  \includegraphics[width=0.96\columnwidth]{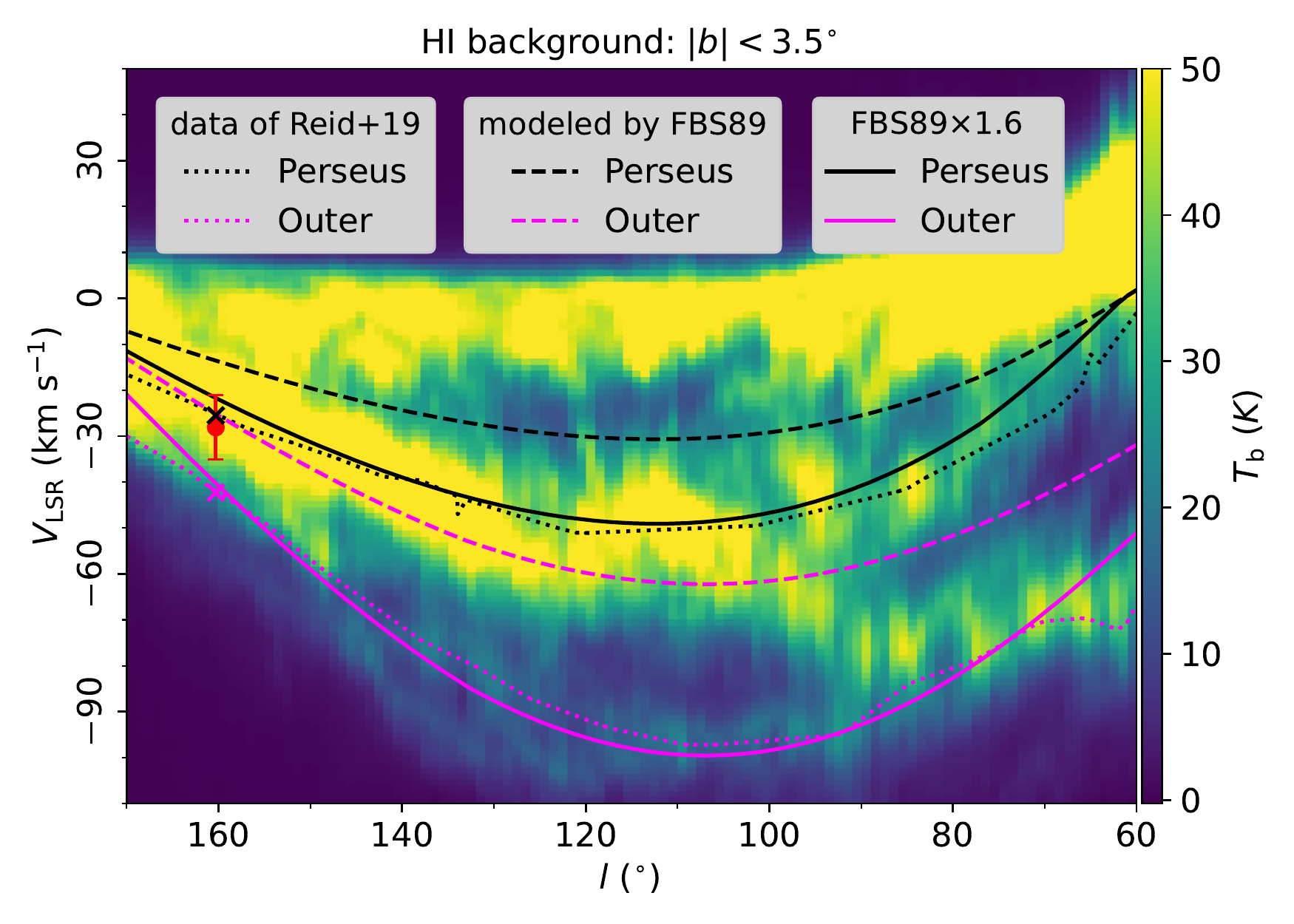}
  \includegraphics[width=0.9\columnwidth]{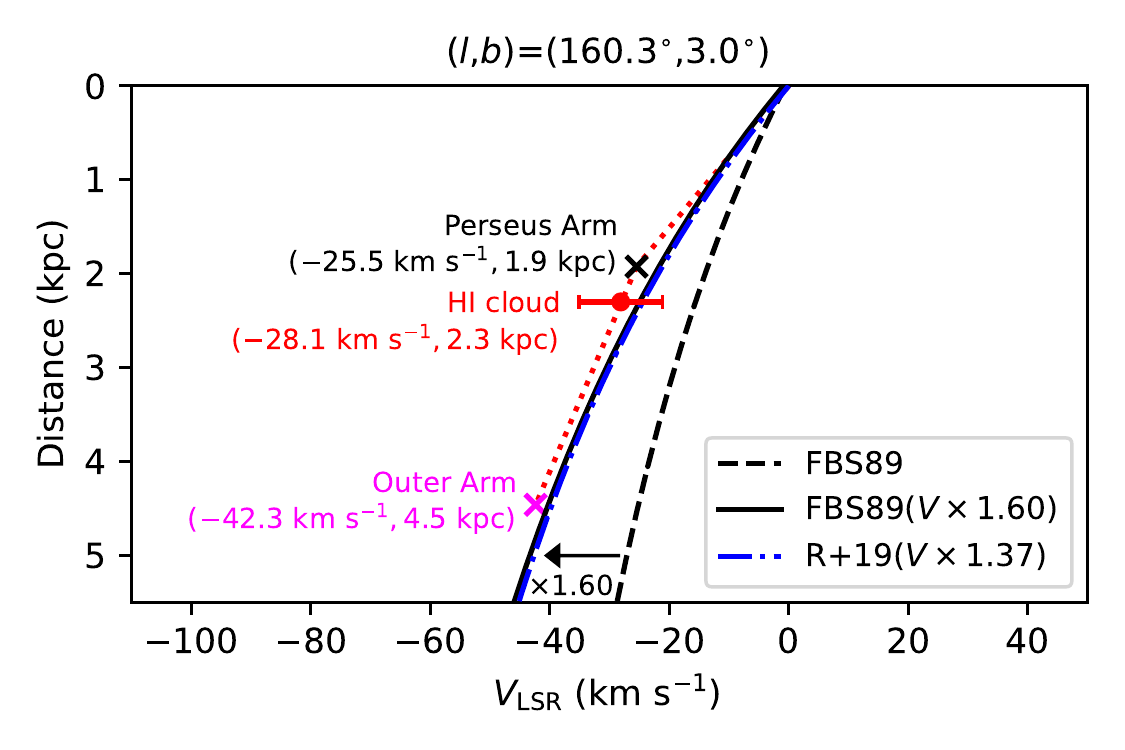}
  \caption{
   {\it Top panel}: The background image is the longitude-velocity plot of H{\tt I} integrated from $|b|<3.5^{\circ}$ of EBHIS \citep{wkf16} in the longitude range of $l$ $=$ 60$^{\circ}$ to 170$^{\circ}$. The dotted lines are the Perseus Arm and the Outer Arm outlined according to \citet{rmb19}. The dashed and solid lines represent the velocity-longitude curves calculated from the original and modified velocity-distance conversions, respectively. At the longitude of the pulsar, the red dot denotes the H{\tt I} absorption of $-28.1~$km~s$^{-1}$ with an error bar of $7~$km~s$^{-1}$, while the two crosses indicate the velocities and distances of Perseus and Outer Arms given by \citet{rmb19}.
  {\it Bottom panel}: Velocity-distance conversion curves in the direction of the pulsar. The dashed black line represents the original conversion from the conventional rotation curve of \citet{fbs89}, while the solid black line has been modified by a multiplicative factor of 1.60 on the negative velocity (see text). The dash-dotted blue line, which lies virtually atop the solid line, represents the modified conversion from the rotation curve of \citet{rmb19}.
  }
  \label{fig:gl-v}
\end{figure}

After the velocities of foreground H{\tt I} clouds are determined, the kinematic distance of the clouds can be estimated based on the Galactic rotation curve and then the distance of a pulsar can be constrained \citep[e.g.][]{wrb80,fw90,wsx+08,vwc+12}. 
The conventional rotation curve that constrains pulsar distances is given by \citet{fbs89} and adopts the International Astronomical Union (IAU) defined parameters of $R_{0}=8.5$~kpc and $\Theta_{0}=220$~$\rm km~s^{-1}$, where $R_{0}$ is the distance to the Galactic Center and $\Theta_{0}$ is the rotation speed of the Galaxy at $R_{0}$ \citep{kl86}. 
Note that the peculiar motions of H{\tt I} clouds in the Perseus Arm and Outer Arm need to be taken into account as they have an additional motion in addition to the circular motion of the rotation curve \citep[see e.g.][]{rmb19,ptm22}, which prevents their distances from being derived directly from the rotation curve. Therefore, it is necessary to modify the velocity-distance conversion in the direction around  $l\sim160^{\circ}$. 
We modify the velocity-distance conversion curve by a factor of 1.6 to the observed radial velocities, similarly to what was done for $100^{\circ}<l<140^{\circ}$ \citep{jrd89,fw90}.
%
Such a modification can be applied to the longitude-distance curve of each spiral arm from a longitude-velocity curve \citep[e.g.][]{v08}, 
so that the newly calculated longitude-velocity curves match the observed spiral arms very well, as shown in in Fig. \ref{fig:gl-v}. 

Particularly, to estimate the distances of the clouds obscuring PSR B0458+46, this modified velocity-distance conversion is used in Fig.~\ref{fig:all}(c). An uncertainty caused by possible non-circular and random peculiar motions of clouds  with a typical value of 7~$\rm km~s^{-1} $ \citep{dl90,rmz09,wwm15} is indicated by the shadowed region in Fig.~\ref{fig:all}(c). 
The nearby cloud for a peak velocity of $V_{\rm LSR} = -7.7$~$\rm km~s^{-1}$ has a distance of $0.6 \pm 0.6$~kpc.
The distant cloud with a peak velocity of $V_{\rm LSR} = -28.1$~$\rm km~s^{-1}$ should be at a distance in the range of [1.9, 3.6]~kpc,  mostly probably 2.7~kpc. Because the pulsar must be behind the distant cloud, 
we therefore adopt the lower distance limit of PSR B0458+46 as being $2.7^{+0.9}_{-0.8}$~kpc. This pulsar is probably located behind the center of the Perseus Arm at 1.9~kpc in this direction (see Fig.~\ref{fig:gl-v}). 

\section{Discussions}

The new distance limit of PSR B0458+46 being $D_{\rm psr}=2.7^{+0.9}_{-0.8}$~kpc has clear implications on the possible association with SNR HB9 and also the Galactic electron density in the outer Galaxy.
Nevertheless, we noticed that \citet{rmb14,rmb19} have measured the parallax-based distances of masers, and developed a new ``universal'' rotation curve for the Milky Way. We discuss if the use of this revised rotation curve can lead to an alternative distance estimate.

\subsection{Alternative distances from an new Galactic rotation curve or simply the H{\tt I} gas location in arms} \label{subsec:reid}

The rotation curve of the Milky Way is fundamental for kinematic distance estimation from the observed radial velocities. Recent years new accurate parallax-based measurements of distances of spiral arm tracers have made significant improvements to the Galactic rotation curve and also $R_0$ and $\Theta_0$ \citep{rmb14,rmb19}. The adoption of this new rotation curve may lead to more reliable velocity-distance conversions. Among several parameter sets in \citet{rmb19}, the A5 fitting is the best with little-assumed priors, and we adopt a set of self-consistent Galactic parameters from the A5 model for the velocity-distance conversions, in which $\Theta_0=236 \pm 7~{\rm km~s^{-1}}$ and $R_0=8.15 \pm 0.15 ~{\rm kpc}$ and $(U,V,W)=(10.6\pm1.2,\ 10.7\pm6.0,\, 7.6\pm0.7)~{\rm km~s^{-1}}$. 

However, the non-uniform rotation of the Milky Way and the presence of streaming motions have to be cautioned for the velocity-distance conversion. As shown in Fig.~\ref{fig:gl-v}, the velocity must be multiplied by a factor of 1.37 in the Galactic longitude range of $l$ $=$ 60$^\circ$ to 170$^\circ$, so that the measured H{\tt I} gas arms can be consistent with the determined Perseus Arm and outer Arm. We found that with such a factor of 1.37, the velocity-distance conversion is almost the same as that of the conventional rotation curve \citep{fbs89} with a factor of 1.60 (see Figure\ref{fig:gl-v}). The resulting distance limit by this new conversion is $3.0 \pm 1.0$~kpc, which in fact is consistent with the result obtained from the conventional rotation curve \citep{fbs89} mainly due to a different factor.

\citet{rdm16} have demonstrated a technique for calibrating kinematic distance based on the positions of spiral arms, although it differs slightly from the method presented here. The association between velocities of known sources and the H{\tt I} absorption lines presents an invaluable tool for distance determination, complementing with parallax-based measurements of spiral arm and the gas kinematic distances derived from Galactic rotation curves. 
As shown in Fig. \ref{fig:gl-v}, the absorption line observed at a velocity of $-28.1\rm~km~s^{-1}$ matches the velocity for the outer side of the Perseus Arm, which directly suggests that the H{\tt I} cloud obscuring PSR B0458+46 is likely associated with the Perseus Arm centered at approximately $1.9 \pm 0.2$ kpc away along the line of sight \citep{xrz06,rmb19}. In fact, a cloud at a distance of $2.3_{-0.7}^{+1.1}$~kpc can be directly derived for the velocity of $-28.1\rm~km~s^{-1}$ from the interpolation of the locations for the Perseus and Outer Arms in this direction as shown Fig. \ref{fig:gl-v}, without involving any rotation curve model, since there are the excellent measurements of the location of two spiral arms \citep{xrz06,rmb19}.

\begin{figure}
  \centering
  \includegraphics[width=0.9\columnwidth]{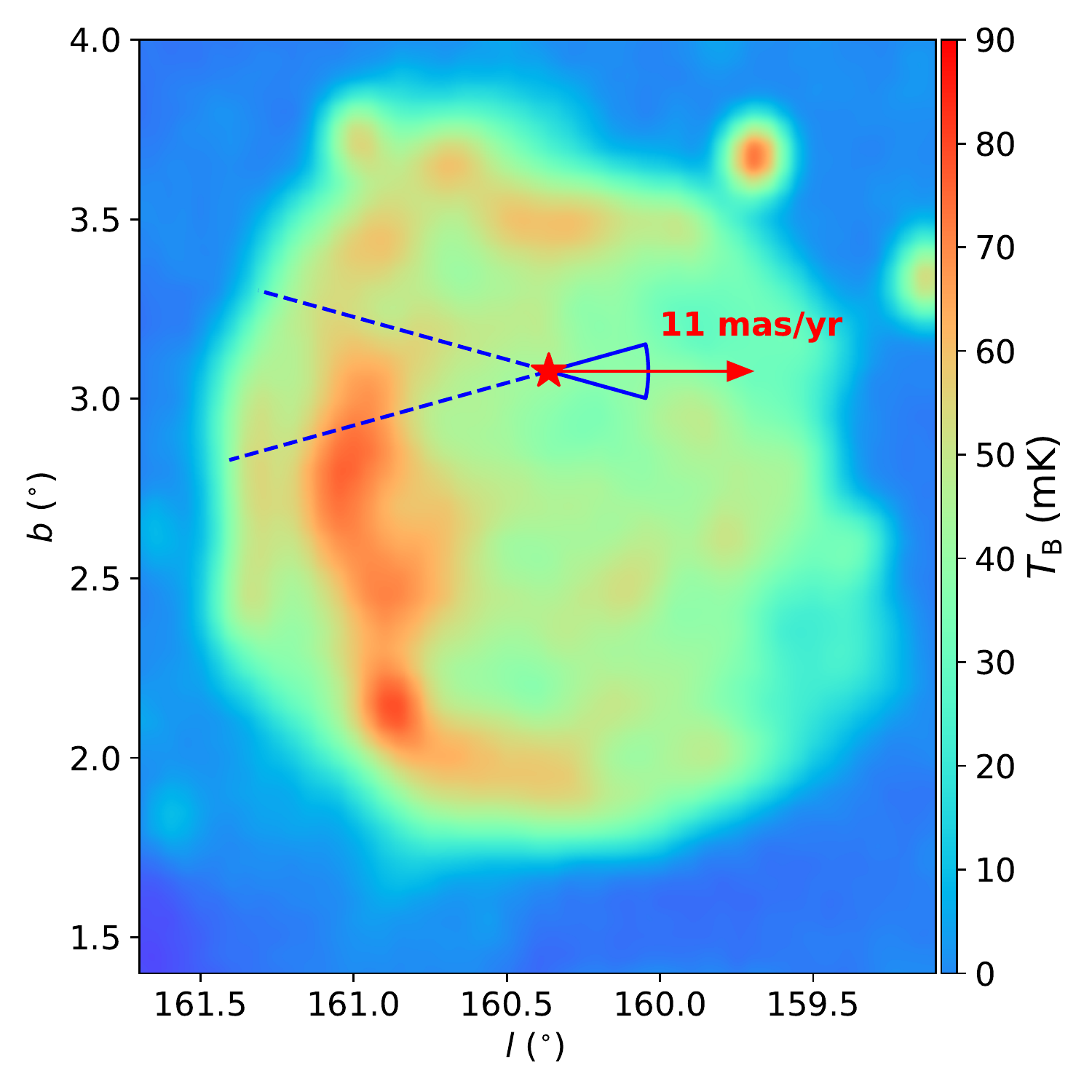}
  \caption{
    PSR B0458+46 (the star) on the radio emission map of the SNR HB9 produced by the 6-cm survey data of \citet{grh+10}. The proper motion of PSR B0458+46 \citep{hla93} is indicated by a solid arrow, with its uncertainty indicated by a blue fan-shaped region and a suggested previous location between dashed lines.
    }
\label{fig:snr}
\end{figure}

\subsection{The association between PSR B0458+46 and SNR HB9} \label{subsec:snr}

PSR B0458+46 is located within the sky area of SNR HB9, leading to a possibility of physical association between the two objects. However, the physical association cannot be determined without verifying the distances of the two objects.

Various methods have been employed to estimate the distance of SNR HB9. 
In the early days, \citet{cl79} derived the relation between the SNR surface brightness $\Sigma$ and the diameter $D$ for a number of SNRs, and got a distance of $1.8$~kpc for SNR HB9. \citet{Milne79} also found a $\Sigma-D$ relation and suggested the distance of $1.3$~kpc for the SNR. Analyzing the H$\alpha$ lines surrounding the SNR, \citet{Lozinskaya81} got an estimated value of $2 \pm 0.8$ kpc derived from the velocity $V_{\rm LSR}$ of $-18 \pm 10$ km$\rm ~s^{-1}$ of H$\alpha$ filaments.  Investigating the early X-ray observation of that SNR, \citet{la95} estimated the distance as 1.5 kpc using the evaporative cloud model. Based on the morphological relevance of the radio image of the SNR and the frequency channel maps in the velocity range of $-3 < V_{\rm LSR} < -9~{\rm km~s^{-1}}$ of H{\tt I} emission in the new H{\tt I} surveys, \citet{lt07} concluded the distance of SNR HB9 is $0.8 \pm 0.4$~kpc. \citet{sey19} analyzed the X-ray, gamma-ray and radio images of the SNR, and also verified the H{\tt I} morphology of different velocities from the new H{\tt I} surveys, and concluded that the distance to the SNR HB9 is $0.6 \pm 0.3$~kpc. \citet{rl22} recently revised it to  $0.7 \pm 0.4$~kpc. By observing the extinction of stars, \citet{zjl20} suggested a distance of $0.54 \pm 0.10~{\rm kpc}$.

Considering the more recent and improved data, it is evident that the distance to SNR HB9 should be less than 1.0 kpc. Thus, the much closer distance of SNR HB9 compared to the larger distance of PSR B0458+46 indicates that they are not physically associated.

Furthermore, the observed proper motion of PSR B0458+46 also does not support the physical association. Its proper motion has been measured by the very long baseline interferometry \citep[VLBI, ][]{hla93}: $\mu_\alpha = -8\pm3$ mas~yr$^{-1}$, $\mu_\delta = 8 \pm 5$~mas~yr$^{-1}$, and timing \citep{hlk04}: $\mu_\alpha\cos(\delta) = 1 \pm 8$~mas~yr$^{-1}$, $\mu_\delta = 42 \pm 19$~ mas~yr$^{-1}$. Because the error bars of the VLBI measurement are much smaller than those of timing, a moving direction and the uncertainty from the VLBI are adopted. The possible region for the birthplace of PSR B0458+46 can be traced back as shown in Fig. \ref{fig:snr}, apparently too far away from the geometric center of the SNR.

\begin{figure}
  \centering
  \includegraphics[width=0.9\columnwidth]{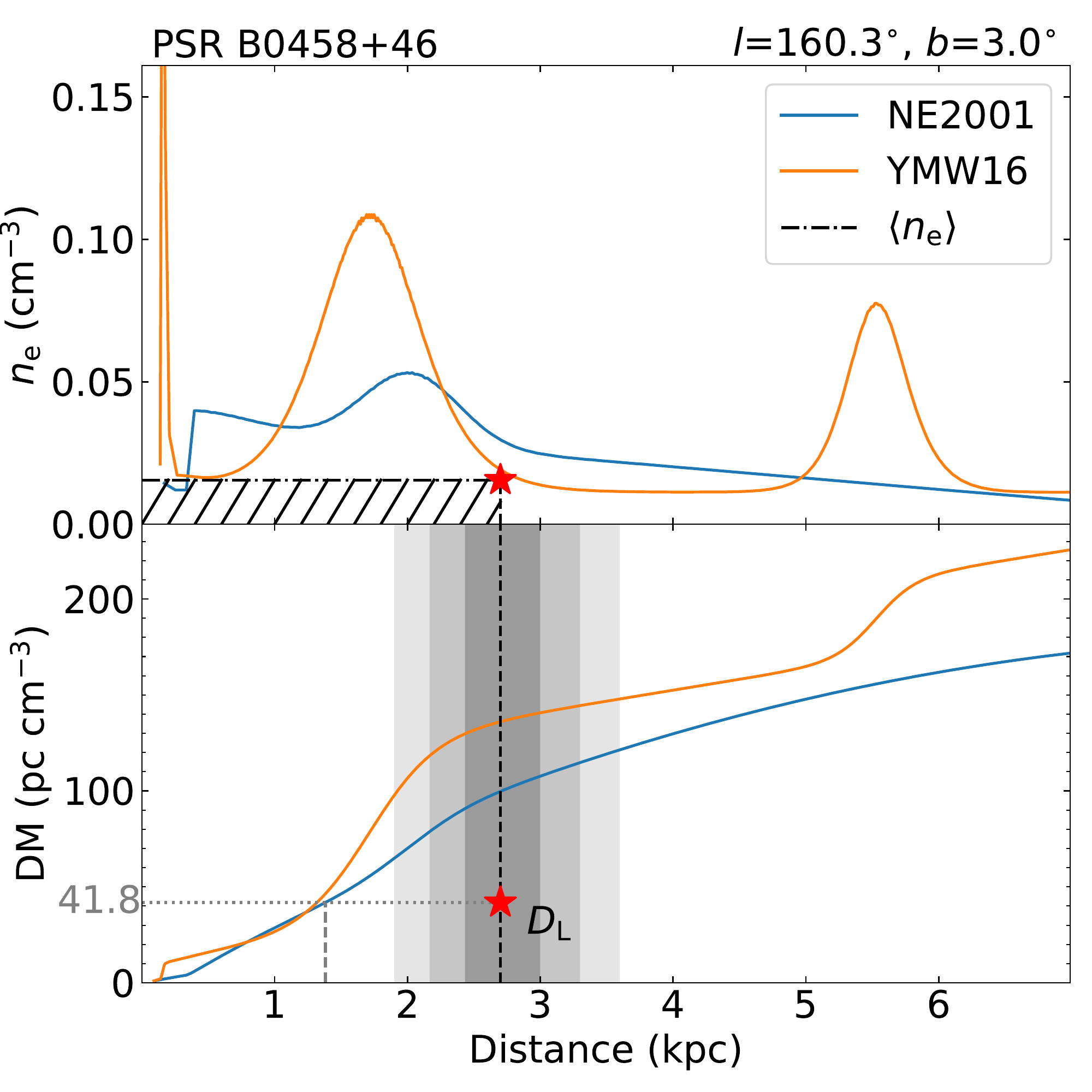}
  \caption{{\it Top panel:} Electron density along the line of sight to PSR B0458+46 in the Galactic electron density models,  NE2001 \citep{cl02} and YMW16 \citep{ymw17}. 
  The area of the hatched rectangle bounded by the lower distance limit and the associated mean electron density must be equal (by definition) to the dispersion measure. The corresponding areas under the two modeled electron density curves are much higher.
  {\it Bottom panel:} Comparison of the DM of PSR B0458+46 with model predictions. The shadowed area stands for possible range of pulsar distance, caused by the random cloud motion of $7~{\rm km~s^{-1}}$.}
\label{fig:electron}
\end{figure}

\subsection{Electron density in the Outer Disk} \label{subsec:electron}

Using a lower distance limit of $D_{\rm L}=$2.7$~{\rm kpc}$ and a DM of 41.834 pc~cm$^{-3}$, the mean electron density between PSR B0458+46 and Earth $\langle n_{\rm e} \rangle$ is determined to be less than ${\rm DM}/D_{\rm L} = 41.834/2700 = 0.0155$~cm$^{-3}$ as depicted in Fig. \ref{fig:electron}. This result is significantly lower than the values previously predicted by models of the Galactic electron density distribution \citep{cl02,ymw17}, indicating the deficiency of electrons in the immediate outer Galaxy in the anti-center region.

To verify the electron density distribution in different parts of our Galaxy, e.g. in the spiral arms and inter-arm regions, in the central bulge and anti-center region, we need to measure the dispersion of pulsed signals and the distance of many pulsars. The latter is challenging to achieve, as demonstrated by this work.

\section{Summary}

By using the FAST, we observe the H{\tt I} absorption lines from the foreground clouds in front of PSR B0458+46. The two absorption lines were detected at velocities of $-7.7$ and $-28.1\rm km~s^{-1}$, respectively. Using the conventional Galactic rotation curve \citep{fbs89} and accounting for the modification on its velocity-distance conversion near the Galactic anti-center \citep{fw90}, the most probable distance of the farthest foreground H{\tt I} cloud in front of this pulsar is around 2.7 kpc.
We try to use the new derived rotation curve of \citet{rmb19} but with a different modification factor to account the streaming motions in the Galactic anti-center, and get a similar distance. In fact, the direct interpolation of the H{\tt I} gas velocities in the parallax-based spiral arms for the absorption clouds with a velocity of  $-28.1\rm km~s^{-1}$ gives a distance of 2.3~kpc. 
we take 2.7~kpc as the lower distance limit of PSR B0458+46 with three reasons:
1) it is consistent with previous standard procedures; 2) it is between the $3.0\pm1.0$~kpc and $2.3^{+1.1}_{-0.7}$~kpc in Section \ref{subsec:reid}; 3) all of them are consistent in the uncertainty.
We therefore conclude that PSR B0458+46 must be located behind the Perseus arm and not physically associated with SNR HB9. The larger distance of PSR B0458+46 implies a much lower electron density in the anti-center region, indicating the deficiency of electrons in the immediate outer Galaxy.
 
Such a new measurement of H{\tt I} absorption lines of pulsars by the sensitive {\it FAST} observations demonstrates the {\it FAST} capability of constraining new kinematic distances of pulsars and contributing to  improve the Galactic electron density distribution model in the near future.

\section*{ACKNOWLEDGEMENTS}
Sincere thanks to the expert referee, Prof. Joel M. Weisberg for the generous, inspiring and kind suggestions and comments. 
FAST is a Chinese national mega-science facility built and operated by the National Astronomical Observatories, Chinese Academy of Sciences. 
This work is supported by the National Natural Science Foundation of China (NSFC, Nos. 11988101 and 11833009) and the Key Research Program of the Chinese Academy of Sciences (Grant No. QYZDJ-SSW-SLH021). In addition,  
TH is supported by the NSFC No. 12003044;
CW is partially supported by NSFC No. U1731120; 
XYG acknowledges the financial support from the 
CAS-NWO cooperation programme (Grant No. GJHZ1865) 
and the NSFC No. U1831103;
LGH thanks the support from the Youth Innovation Promotion Association CAS;
DJZ is supported by the Cultivation Project for the {\it FAST} scientific Payoff and Research Achievement of CAMS-CAS;
JX is partially supported by NSFC No. U2031115 and the National SKA program of China (Grant No. 2022SKA0120103).


In the data processing, the python packages \texttt{scipy} \citep{scipy} and \texttt{numpy} \citep{numpy} are employeed. 

\section*{DATA AVAILABILITY}
Original {\it FAST} observation data can be accessible in the {\it FAST} Data Center one year after observations.
All processed data as plotted in this paper, including these in appendix, can be obtained from the authors with a kind request.
The pulsar profile data are available for \url{http://zmtt.bao.ac.cn/psr-fast/}.

\bibliographystyle{mnras}
\bibliography{fastHIabs} 

\appendix


\section{The Baseline and Standing-wave Fitting and a High-velocity H{\tt I} Cloud}
\label{app:hi}

\begin{figure}
  \centering
  \includegraphics[width=0.95\columnwidth]{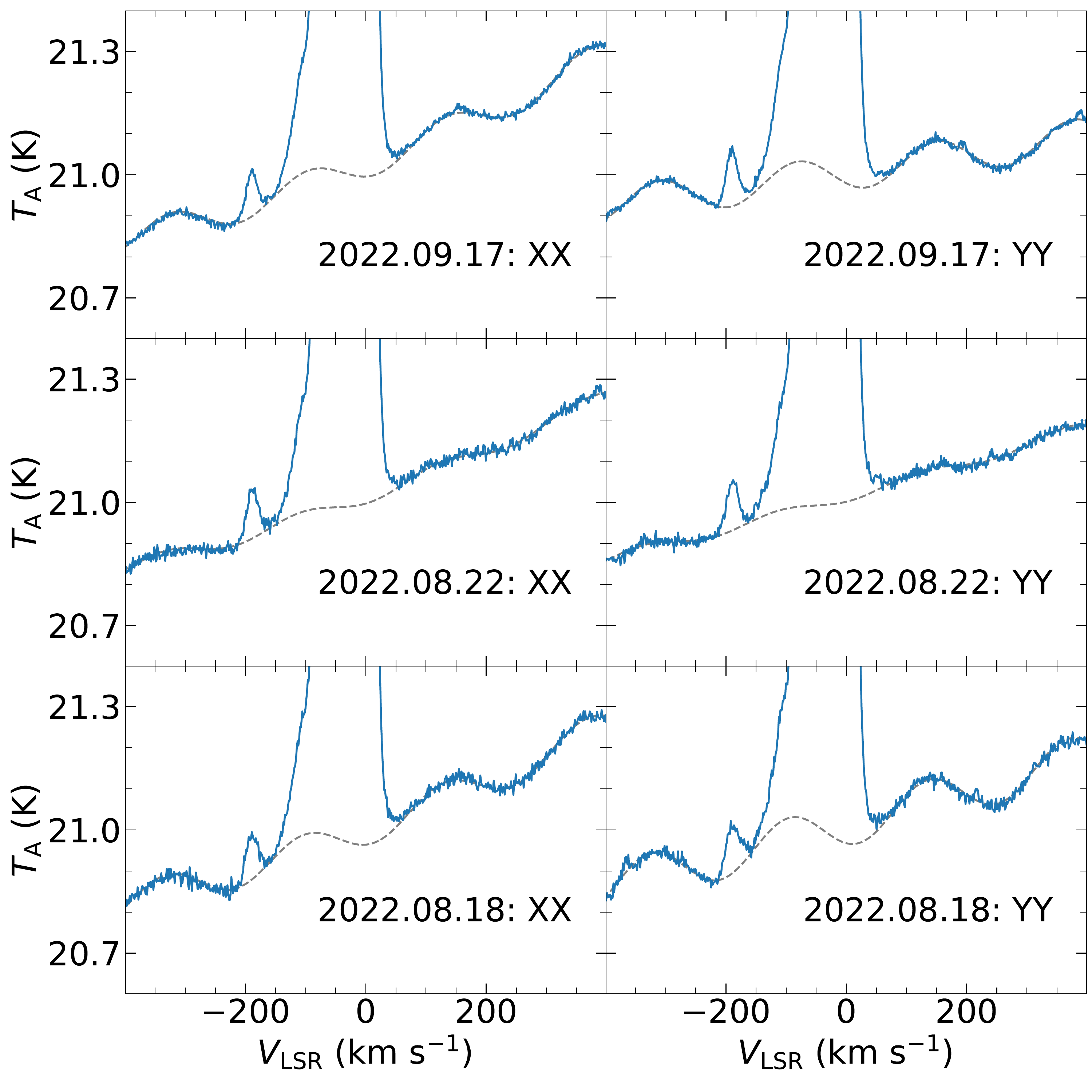}
  \includegraphics[width=0.95\columnwidth]{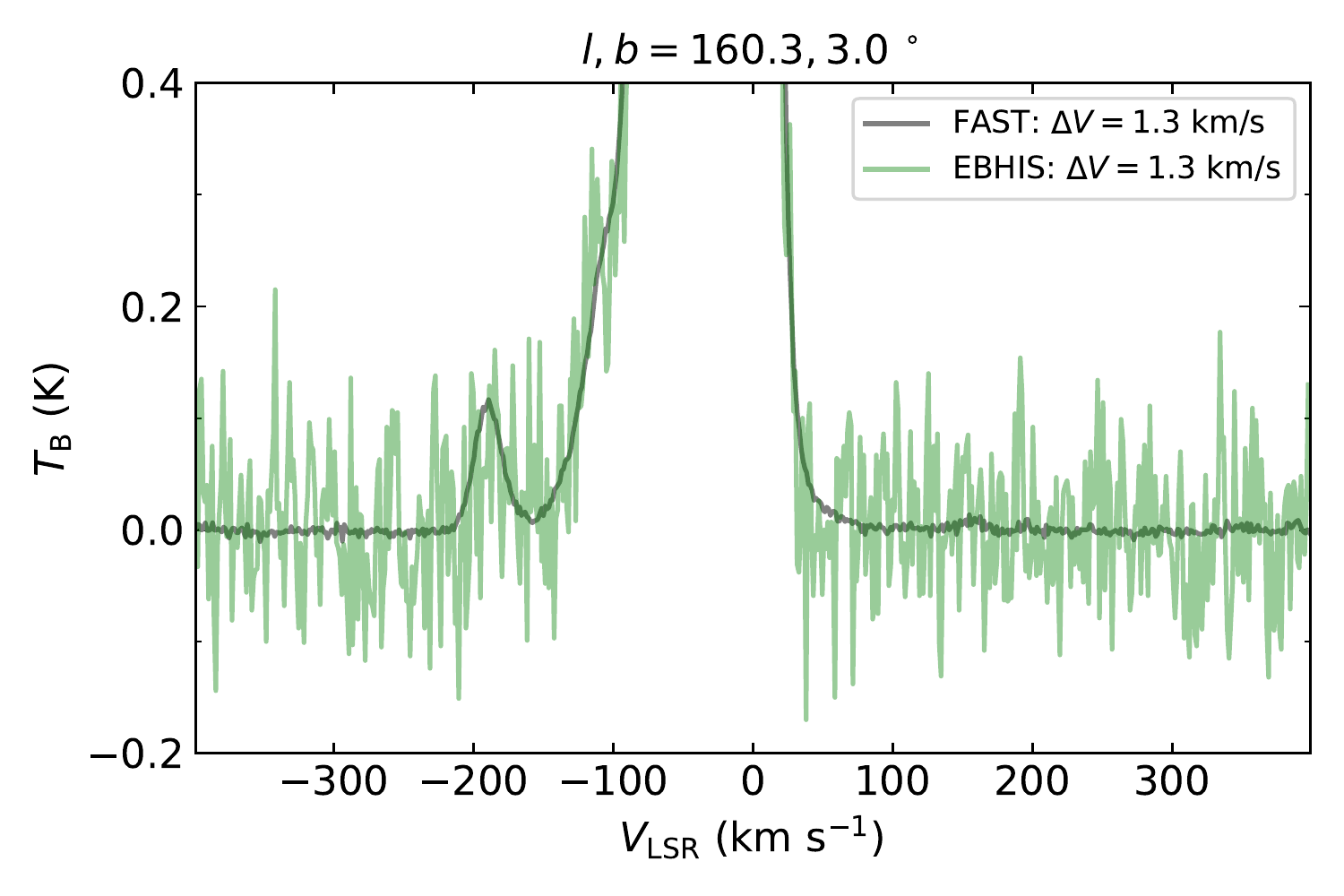}
  \caption{
    {\it FAST} detected H{\tt I} emission lines in the resolution of 1.3 km~s$^{-1}$ with and without subtraction of the standing waves and baselines,  zoomed to show the details.
    In the {\it top panels}, the sinusoidal-like baselines on the H{\tt I} spectra are shown for XX and YY polarization in three observation sessions, respectively. The baseline-removed and summed result is shown in the {\it bottom panel} and compared with that from the EBHIS \citep{wkf16}. A high-velocity H{\tt I} cloud is detected at the velocity of around $-$190 km~s$^{-1}$ by the {\it FAST} observation.}
\label{fig:hi_tiny}
\end{figure}

There is a new H{\tt I} emission line at the velocity of $-190$ km~s$^{-1}$ clearly detected by {\it FAST} observations, as shown in the bottom panel of Fig. \ref{fig:hi_tiny}. Data from the EBHIS are shown together for comparison.

To verify the detection, we present and check the raw spectra from XX and YY polarization products of all three observation sessions in the top of Fig. \ref{fig:hi_tiny}. Though the standing waves and baselines are different for each session, the H{\tt I} line for this high-velocity H{\tt I} cloud presents in all spectra. Therefore, the cloud revealed by this weak high-velocity line is a true detection.

In general, the standing waves and baseline can be removed by a model-fitting to the line-off region of the pulse-off spectrum with:
\begin{equation}
\label{eq:stand}
    T_{\rm A}(V_{\rm LSR}) = T_{\rm SW} \sin(\omega_0 V_{\rm LSR}+\omega_1 V_{\rm LSR}^2 + \phi) + k \cdot V_{\rm LSR} + T_0 ,
\end{equation}
here the parameters $k$ and $T_0$ are set for the linear term, and $T_{\rm SW}$, $\omega_0$, $\omega_1$ and $\phi$ are set for the waves along $V_{\rm LSR}$. 

This newly detected emission line should come from a distant or high-velocity H{\tt I} cloud \citep[as reviewed by][]{ww97}, similar with the clouds pointed out in the Appendix D of \citet{smm22}. 
Such a cloud can be investigated by the piggy-back recorded spectral data during the {\it FAST} GPPS survey \citep{hww21}, as done by \citet{Hong22}. 

\section{Detection of pulsar H{\tt I} Absorption in three sessions}
\label{app:psr}

\begin{figure}
  \centering
  \includegraphics[width=0.9\columnwidth]{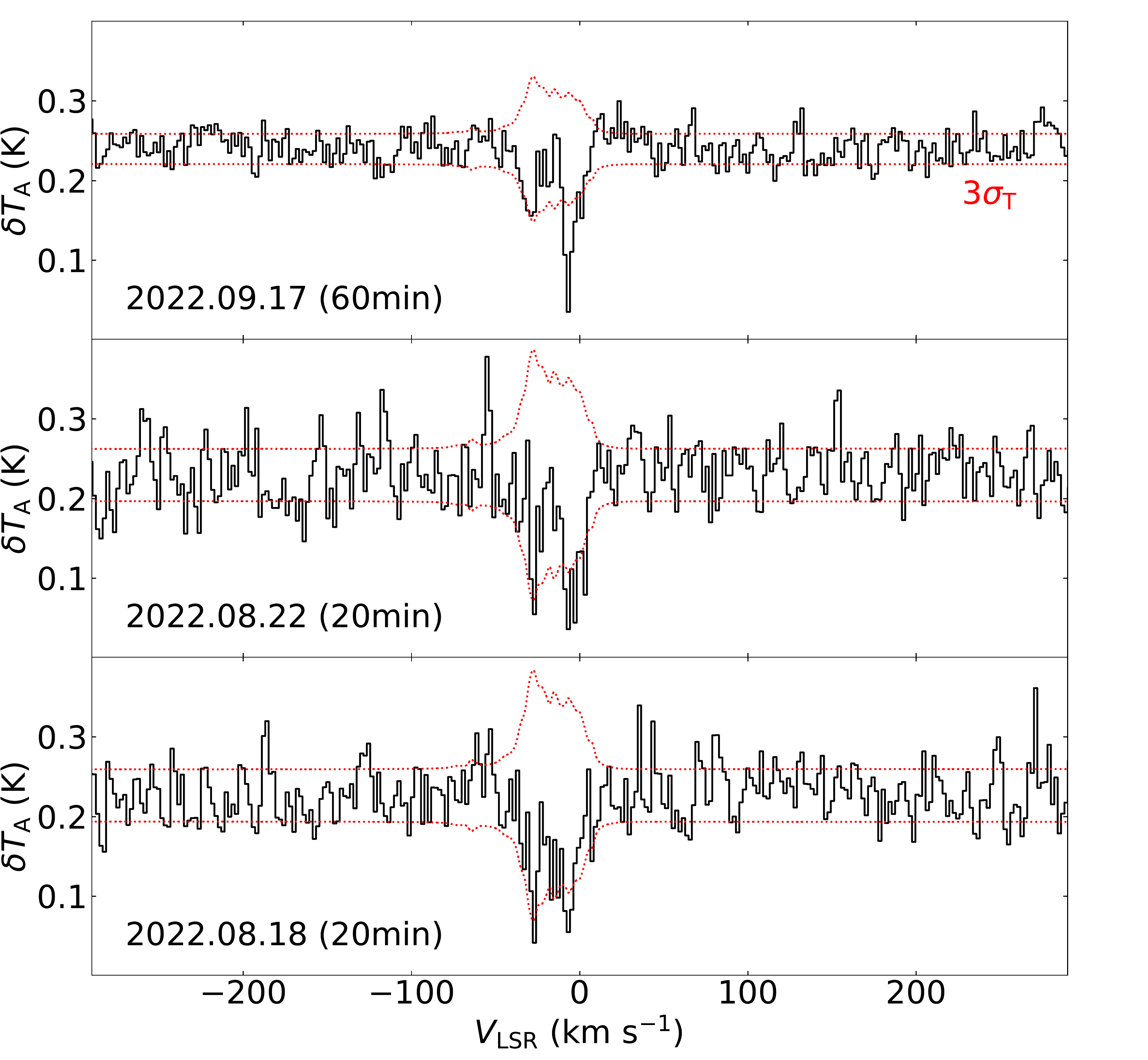}
  \caption{
    Pulsar spectra obtained in three observational sessions, showing the absorption lines. 
    The thin dot line represents the fluctuations caused by the system temperature and extra temperature from H{\tt I} emission lines.
    }
\label{fig:ghost}
\end{figure}

We present the detected absorption in three observation sessions in Fig. \ref{fig:ghost}.
All three pulsar spectra show the two absorption lines clearly.
Due to the strength of the H{\tt I} line, the radiometer noise in the on-HI-line portion of a spectrum is enhanced with respect to the noise in the off-line portion of the same spectrum \citep{wsx+08}. The noise, in the temperature units, at any frequency $f$, is given by
\begin{equation} \label{equ:temp}
    \sigma_{\rm T}(f) = \frac{T_{\rm sys}+T_{\rm HI}(f)}
    {\sqrt{n_{\rm p}t_{\rm int}\Delta{f}}},
\end{equation}
where $T_{\rm sys}$ is measured system temperature, $T_{\rm HI}$ is H{\tt I} temperature, $n_{\rm p}=2$ is the number of the polarization probes, $t_{\rm int}$ is the observational time of the pulse-on spectra, and $\Delta{f}$ is the frequency resolution of the pulsar spectra. 
The system temperature is $T_{\rm sys} \simeq 20$~K \citep{jth20}, and the temperature of H{\tt I} lines $T_{\rm HI}(f)$ shown in Fig.~\ref{fig:all}(a) is about 40 -- 70~K, and then fluctuations due to ``system'' temperature are therefore enlarged as shown in 
 Fig.~\ref{fig:ghost} and also Fig.~\ref{fig:all}(b).

\bsp	
\label{lastpage}
\end{document}